\documentclass{aa}
\usepackage{amssymb,alltt}
\usepackage{latexsym}
\usepackage{graphicx}
\graphicspath{{PS/}}
\usepackage{psfig}
\psfigurepath{./PS:.}
\usepackage[english]{babel}
\usepackage{indentfirst}
\usepackage{psfig}
\usepackage{natbib}

\psfigurepath{./../PS}
\vspace{1cm}
\newcommand{\be}{\begin{eqnarray}}
\newcommand{\ee}{\end{eqnarray}}

\newcommand{\bitem}{\begin{itemize}}
\newcommand{\eitem}{\end{itemize}}

\newcommand{\wave}{{\cal W}}

\newcommand{\tkle}{${\tilde \kappa}_l^{(E)}$}
\newcommand{\tklb}{${\tilde \kappa}_l^{(B)}$}

\begin{document}
\title{Weak Lensing Mass Reconstruction \\ using Wavelets}
\author{Jean-Luc Starck, Sandrine Pires and Alexandre R\'efr\'egier}

\author{Jean-Luc Starck \and Sandrine Pires  \and Alexandre R\'efr\'egier}
\institute{DAPNIA/SEDI-SAP, Service d'Astrophysique, CEA-Saclay, F-91191 Gif-sur-Yvette Cedex, France}
 

\offprints{jstarck@cea.fr}
 
\date{\today}

 

\abstract{This paper presents a new method for the reconstruction of
weak lensing mass maps. It uses the multiscale entropy concept, which
is based on wavelets, and the False Discovery Rate (FDR) which allows
us to derive robust detection levels in wavelet space. We show that
this new restoration approach outperforms several standard techniques
currently used for weak shear mass reconstruction. This method can
also be used to separate E and B modes in the shear field, and thus
test for the presence of residual systematic effects. We concentrate
on large blind cosmic shear surveys, and illustrate our results using
simulated shear maps derived from N-Body $\Lambda$CDM simulations
\citep{vale03} with added noise corresponding to both
ground-based and space-based observations.}

\maketitle 
\markboth{Weak Lensing Mass Reconstruction}{}


\section{Introduction}

\indent Weak Gravitational Lensing provides a unique method to map
directly the distribution of dark matter in the universe  
\citep{bartelmann99,mellier99,Waerbeke01,mellier02,refregier03}. This
method is based on the weak distortions that lensing induces in the
images of background galaxies as light travels through intervening
structures. This method is now widely used to map the mass of clusters
and superclusters of galaxies and to measure the statistics of the
cosmic shear field on large scales.\\

\indent Ongoing efforts are made to improve the detection of cosmic
shear on existing telescopes and future instruments dedicated to
survey cosmic shear are planned. Several methods are used to derive
the lensing shear from the shapes of background galaxies. But the
shear measurements obtained are always noisy, and when it is converted into a
map of the projected mass $\kappa$, the result is dominated by the
noise.\\

\indent Several methods have been devised to reconstruct the projected
mass distribution from the observed shear field. The first
non-parametric mass reconstruction was proposed by \citet{kaiser93} and
further improved by
\citet{bartelmann95,kaiser95,schneider95,squires96}. These methods are
based on linear inversion methods based on smoothing with a fixed
kernel. Non-linear reconstruction methods were proposed using a
maximum likelihood approach \citep{squires96,bartelmann96,Seitz98} or
using the maximum entropy method \citep{bridle98,marshall02}. \\
\indent In this paper, we describe a method for weak lensing mass
reconstruction based on a wavelet decomposition. We use an iterative
filtering method with a multiscale entropy regularisation to filter
the noise. We discuss how this decomposition and regularisation
functional is particularly well adapted to this problem. In the
process, we identify significant wavelet coefficients using the False
Discovery Rate method \citep{miller01,hopkins02} and show how this is
superior to the standard $n \sigma$ thresholding. The FDR method
adapts its threshold to the features of the data.  We concentrate on
large blind cosmic shear surveys and use the ray-tracing simulations
of \citet{vale03} to test our results. We compare the performance of
our method to Gaussian and Wiener filtering for the reconstruction of
the mass field in these simulations. We consider conditions similar to
both ground-based and space-based cosmic shear surveys. We also
discuss how our method differs from other methods based on the maximum
entropy prior.\\

\indent In section~\ref{sect_wsmr}, we present the weak shear mass
reconstruction problem. The earlier methods which have been proposed
to reconstruct the mass map are described in
section~\ref{earlier}.  Section~\ref{sect_mult_entrop} presents
the Multiscale Entropy method and explains why it is a good
alternative to standard methods. We also propose a modification of the
Multiscale Entropy, we use the False Discovery Rate (FDR) method for
detecting the significant wavelet coefficients. A set of experiments
designed to test our method are described
section~\ref{sect_results}. Our conclusions are summarized in section
~\ref{sect_conclusion}.

\section{Weak lensing mass reconstruction}
\label{sect_wsmr}
\subsection{Weak lensing}
In weak lensing surveys, the shear $\gamma_i({\mathbf \theta})$ with
$i=1,2$ is derived from the shapes of galaxies at positions ${\mathbf
\theta}$ in the image. The shear field $\gamma_i({\mathbf \theta})$
can be written in terms of the lensing potential $\psi({\mathbf
\theta})$ as (see eg. \citet{bartelmann99})
\begin{eqnarray}
\label{eq:gamma_psi} 
\gamma_1 & = & \frac{1}{2}\left( \partial_1^2 -
\partial_2^2 \right) \psi \nonumber \\ \gamma_2 & = & \partial_1
\partial_2 \psi,
\end{eqnarray}
where the partial derivatives $\partial_i$ are with respect to
$\theta_i$.  The convergence $\kappa({\mathbf \theta})$ can also be
expressed in terms of the lensing potential as
\begin{equation}
\label{eq:kappa_psi}
\kappa =  \frac{1}{2}\left(\partial_1^2 + \partial_2^2 \right) \psi
\end{equation}
and is related to the surface density $\Sigma({\mathbf \theta})$
projected along the line of sight by
\begin{equation}
\kappa({\mathbf \theta})=\frac{\Sigma({\mathbf \theta})}{\Sigma_{\rm crit}}
\end{equation}
where the critical surface density is given by
\begin{equation}
\Sigma_{\rm crit}=\frac{c^2}{4\pi G} \frac{D_{s}}{D_{l}D_{ls}}
\end{equation}
and $G$ is Newton's constant, $c$ is the speed of light and $D_{s}$,
$D_{l}$ and $D_{ls}$ are the angular-diameter distances between the
observer and the galaxies, the observer and the lens, and the lens and
the galaxies. In practice, the galaxies are not at a fixed redshift,
and the expression for $\kappa$ is an average of the redshift of the
galaxies (see eg. \citet{bartelmann95}). The lensing effect is said to
be weak or strong if $\kappa \ll 1$ or $\kappa \gtrsim 1$,
respectively.

The left panel of Fig.~\ref{kb0} shows a simulated convergence map
derived from ray-tracing through N-body cosmological simulations
performed by \citet{vale03}. The cosmological model is taken to be a
concordance $\Lambda$CDM model with parameters $\Omega_M=0.3$,
$\Omega_\Lambda=0.7$, $h=0.7$ and $\sigma_8=0.8$. The simulation
contains $512^3$ particles with a box size of $300 h^{-1}$ Mpc. The
resulting convergence map covers $2 \times 2$ degrees with $1024
\times 1024$ pixels and a assume a galaxy redshift of 1. The
overdensities correspond to the haloes of groups and clusters of
galaxies. The rms value of $\kappa$ binned in 0.12 arcmin pixels is
$\sigma_\kappa = 0.023$. The typical values of $\kappa$ are thus of
the order of a few percent, apart from the core of massive halos (see
figure~\ref{kb0}). The weak lensing condition therefore holds in most
regions of the sky and will be assumed throughout this paper.

\subsection{Mass inversion}

The weak lensing mass inversion problem consists of reconstructing the
projected (normalized) mass distribution $\kappa({\mathbf \theta})$
from the measured shear field $\gamma_i({\mathbf \theta})$ by
inverting equations (\ref{eq:gamma_psi}) and (\ref{eq:kappa_psi}). 
(Magnification information can also be used to improve the reconstruction [see \citet{bridle98}], 
but is typically more noisy than the shear measurements and has not been considered
in this paper). For this purpose, we take the Fourier transform of these equations and
obtain 
\begin{equation}
\hat{\gamma_i} = \hat{P_i} \hat{\kappa},~~~i=1,2
\end{equation}
where the hat symbol denotes Fourier transforms and we have defined
$k^2 \equiv k_1^2 + k_2^2$ and
\begin{eqnarray}
\hat{P_1}(\mathbf k) & = & \frac{k_1^2 - k_2^2}{k^2} \nonumber \\
\hat{P_2}(\mathbf k) & = & \frac{2 k_1 k_2}{k^2},
\end{eqnarray}
with $\hat{P_1}(k_1,k_2) \equiv 0$ when $k_1^2 = k_2^2$, and
$\hat{P_2}(k_1,k_2) \equiv 0$ when $k_1 = 0$ or $k_2 = 0$. 

\begin{figure*} 
\centerline{
\hbox{
}}
\caption{Left: simulated convergence map from (Vale and White, 2003) for a
$\Lambda$CDM model. The region shown is $2\times2$ square
degree. Right: Shear map superimposed on the convergence map , and
right shear map. The size and direction of each line gives the
amplitude and position angle of the shear at this location on the
sky.}
\label{kb0}
\end{figure*}

The shear map $\gamma_i$ can be calculated from the convergence map
$\kappa$ using these expressions.  The right panel of Fig. \ref{kb0},
shows the shear field associated with the simulated convergence
field. As is customary, the direction and size of the line segment
represent the orientation and amplitude of the shear. The rms shear in
the 0.12 amin pixels of the resulting map is $\sigma_{\gamma} =
\langle \gamma_1^2+\gamma_2^2 \rangle^{\frac{1}{2}} \simeq 0.023$.
 
Note that to recover $\kappa$ from $\gamma_1$ (resp. $\gamma_2$),
there is a degeneracy when $k_1^2 = k_2^2$ (resp. when $k_1 = 0$ or
$k_2 = 0$). To recover $\kappa$ from both $\gamma_1$ and $\gamma_2$,
there is a degeneracy only when $k_1 = k_2 = 0$. Therefore, the mean
value of $\kappa$ cannot be recovered from the shear maps. This is a
special instance of the well known mass-sheet degeneracy in the weak
lensing reconstruction if only shear information is available (see
eg.\citet{bartelmann95} for a discussion).


In practice, the observed shear $\gamma_i$ is obtained by averaging
over a finite number of galaxies and is therefore noisy. The 
relations between the observed data $\gamma_{1b},\gamma_{2b}$ binned in pixels
of area $A$ and the true mass map $\kappa$ are given by:
\begin{equation}
\label{eq_gamma}
\gamma_{ib} = P_i * \kappa + N_i
\end{equation}
where $N_1$ and $N_2$ are noise contributions with zero mean and
standard deviation $\sigma_n \simeq \sigma_\epsilon/\sqrt{N_g}$, where
$N_g = n_g A$ is the average number of galaxies in a pixel and $n_g$
is the average number of galaxies per arcmin$^2$. The rms shear
dispersion per galaxy $\sigma_\epsilon$ arises both from measurement
errors and the intrinsic shape dispersion of galaxies. In this
analysis, we will assume $\sigma_\epsilon \simeq 0.3$ as is
approximately found for ground-based and space-based weak lensing
surveys. Typical values for the surface density of usable galaxies for
weak lensing are
\begin{itemize}
\item $n_g= 20$  gal/arcmin$^2$ for ground-based surveys.
\item $n_g= 100$ gal/arcmin$^2$ for space-based surveys.
\end{itemize}
From the central limit theorem, this means that for pixels with $A \gtrsim 1$
amin$^{2}$, the noise $N_i$ is, to a good approximation, Gaussian in both cases
and is uncorrelated (see \citet{marshall02} for a direct treatment of individual galaxy 
shears using the MEM method).

\subsection{The Inverse Filter: E and B mode}

We can easily derive an estimation of the mass map by inverse
filtering by noticing that
\begin{equation}
\label{eq:p1p2}
\hat{P_1}^2+\hat{P_2}^2=1.
\end{equation}
The least square estimator $\hat{\tilde \kappa}_l^{(E)}$ of the
convergence $\hat{\kappa}$ in the Fourier domain is:
\begin{eqnarray}
\hat{\tilde \kappa}_l^{(E)} & = & \hat{P_1} \hat{\gamma}_{1b}+
  \hat{P_2}\hat{\gamma}_{2b} 
\label{eqn_reckE}
\end{eqnarray}
The relation between this estimator and the true mass map is 
$\hat{\tilde \kappa}_l^{(E)} = \hat{\kappa} + \hat{N}$,
where $\hat{N} = \hat{P_1} \hat{N_1} + \hat{P_2} \hat{N_2}$.

Just as any vector field, the shear field $\gamma_i({\mathbf \theta})$
can be decomposed into a gradient, or electric ($E$), component, and a
curl, or magnetic ($B$), component. Because the weak lensing arises
from a scalar potential (the Newtonian potential), it can be shown
that weak lensing only produces $E$-modes. On the other hand, residual
systematics arising from imperfect correction of the instrumental PSF
or telescope aberrations, generally generates both $E$ and $B$
modes. The presence of $B$-modes is thus used to test for the presence
of residual systematic effects in current weak lensing surveys. 
\label{eb}

The decomposition of the shear field into each of these components can
be easily performed by noticing that a pure $E$-mode can be
transformed into a pure $B$ mode by a rotation of the shear by
45$^{\circ}$: $\gamma_{1} \rightarrow - \gamma_{2}$, $\gamma_{2}
\rightarrow \gamma_{1}$. As a result, we can form the
following estimator for the $B$-mode ``convergence'' field
\begin{equation}
\hat{\tilde \kappa}_l^{(B)} = \hat{P_2} \hat{\gamma_{1b}} - \hat{P_1} *
\hat{\gamma_{2b}},
\end{equation}
and check that it is consistent with zero in the absence of systematics.



\begin{figure}[htb]
 \centerline{
\hbox{
}}
\caption{In the previous region of $2\times2$ square degrees, noisy mass map $\kappa^{(E)}_l$ for the same simulation with $n_g=100$ gal/arcmin$^{2}$, corresponding to space-based
observations. Even in this case, the unfiltered mass map is dominated by noise.}
\label{kappab}
\end{figure}

As follows from equation~(\ref{eq:p1p2}), the noise $N^{(E)}$ and
$N^{(B)}$ in \tkle and \tklb is still Gaussian and uncorrelated.  The
inverse filtering does not amplify the noise, but \tkle and \tklb may
be dominated by the noise if $N^{(E)}$ and $N^{(B)}$ are large, which
is the case in practice. Fig.~\ref{kappab} shows the reconstructed
mass map using equation~\ref{eqn_reckE} when a realistic Gaussian
noise has been added to the shear maps plotted in Fig.~\ref{kb0}
right.  As expected, it is dominated by noise.  This has motivated the
development of different methods in the past which we describe below.

\section{Earlier Mass Inversion Methods}
\label{earlier}
\subsection{Linear Filtering}

\begin{figure*}[htp]
 \centerline{
\hbox{
\hspace{0.2cm}
}}
\caption{Reconstruction error as a function of the kernel size
$\sigma_G$ (in 0.12 amin pixels) for the Gaussian smoothing method,
with $n_g=20$ gal/amin$^{2}$ (left) and $n_g=100$ gal/amin$^{2}$ (right).}
\label{sig}
\end{figure*}
 
The standard method \citep{kaiser93} consists in convolving
the noisy mass map \tkle with a Gaussian window $G$ with standard
deviation $\sigma_G$:
\begin{equation}
{\tilde \kappa}_G^{(E)} =  G * {\tilde \kappa}_l^{(E)}
                        =   G * P_1 * \gamma_{1b} + G * P_2 * \gamma_{2b}
\end{equation}
\indent The quality of the resulting estimation depends strongly on
the value of $\sigma_G$.  Fig. \ref{sig} shows the variation of the
error between the original mass map
$\kappa$ shown in Fig.\ref{kb0} and the filtered mass map ${\tilde
\kappa}_G^{(E)}$. For this simulation, the optimal value of $\sigma_G$
lies between 5 and 10 pixels (1 pixel = 0.12 arcmin) for space
observations (i.e. $n_g=100$ gal/amin$^2$) and lies between 20 and 25 pixels for ground observations 
( i.e. $n_g=20$ gal/amin$^2$).

An alternative to Gaussian filtering is the Wiener filtering, obtained
by assigning the following weight to each $k$-mode
\begin{equation}
w(k)=\frac{|\hat{\kappa}(k)|^2}{|\hat{\kappa}(k)|^2+|\hat{N}(k)|^2}
\end{equation}
Where $|\hat{\kappa}(k)|^2$ is a model of the true convergence power
spectrum and is in practice derived from the data. Wiener filtering is
known to be optimal when both the signal and the noise are a
realization of a Gaussian Random Field. As can be seen from
Fig.\ref{kb0}, this assumption is not valid for weak lensing mass maps
which display non-Gaussian features such as galaxy clusters, groups
and filaments.  Even in this case, Wiener Filtering nevertheless leads
to reasonable results, generally better than the simple Gaussian
filtering.

\subsection{Maximum Entropy Method}
The Maximum Entropy Method (MEM) is well-known and widely used in
image analysis in astronomy (see \citet{bridle98,starck:sta01_1,marshall02,starck:book02}
for a full description).  It considers both the data and the solution
as probability density functions and find the solution using a
Bayesian approach and adding a prior (the entropy) on the
solution. Several definitions of entropy exists. The most common is
the definition proposed in \citet{entropy:gull91}:
\begin{eqnarray}
 H_g(\kappa)  = \sum_{x}\sum_{y} \kappa(x,y) - m(x,y)
  -  \kappa(x,y)  \ln\left(\frac{\kappa(x,y)}{m(x,y)}\right) \nonumber
\end{eqnarray}
where $m$ is a model, chosen typically to be a sky background.
$H_g$ has a global maximum at $\kappa=m$. MEM
does not allow negative values in the solution, which is unnatural for wide
field weak shear data or the CMB data, where we measure
fluctuations around zero. To overcome this, it has been
proposed to replace $H_g$ \citep{entropy:hobson04} by:
\begin{eqnarray}
 H_{+/-}(\kappa) & = & \sum_{x} \sum_{y} \psi(x,y) - 2m - \nonumber \\
                 &    &  - \kappa(x,y) \ln\left(\frac{\psi(x,y) + \kappa(x,y)}{2m}\right)
\end{eqnarray}
where $\psi(x,y) = \sqrt{\kappa^2(x,y) + 4m^2}$. Here $m$ does not play the same role. It is a constant fixed to
the expected signal rms.

More generally MEM method presents many drawbacks
\citep{entropy:narrayan86,starck:sta01_1} and various refinements of
MEM have been proposed over the years
\citep{entropy:weir92,entropy:bontekoe94,starck:pan96,starck:sta01_1}.
The last developments have lead to the so called Multiscale Entropy
\citep{starck:pan96,starck:sta01_1,entropy:hobson04} which is based on
an undecimated isotropic wavelet transform (\`a trous algorithm)
\citep{starck:book98}.  It has been shown that the main MEM drawbacks
(model dependent solution, oversmoothing of compact objects, \dots)
disappear in the wavelet framework.  A full discussion and comparison
between different restoration methods can be found in
\citet{starck:sta02_2}.

\section{Multiscale Entropy Restoration}
\label{sect_mult_entrop}

\subsection{The Multiscale Entropy}
The Undecimated Isotropic Wavelet Transform (UIWT)  
decomposes an $n \times n$ image $I$ as a superposition of
the form
\[
I(k,l) = {c_J}_{k,l} + \sum_{j=1}^{J} w_{j,k.l},
\]
where $c_{J}$ is a coarse or smooth version of the original image $I$
and $w_j$ represents the details of $I$ at scale $2^{-j}$ (see Starck
et al.\citep*{starck:book98,starck:book02} for details).  Thus, the
algorithm outputs $J+1$ sub-band arrays of size $n \times n$. We will
use an indexing convention such that $j = 1$ corresponds to the finest
scale (high frequencies).  The Multiscale Entropy concept
\citep{starck:pan96} consists in replacing the standard MEM prior
(i.e. the Gull and Skilling entropy) by a wavelet based prior. The
entropy is now defined as
\begin{equation}
 H(I) = \sum_{j=1}^{J-1} \sum_{k,l} h(w_{j,k.l}).
\end{equation}
In this approach, the information content of an image is viewed as
sum of information at different scales. The function $h$ defines the
amount of information relative to a given wavelet coefficient.
Several functions have been proposed for $h$:
\begin{itemize}
\item{LOG-MSE:} The Multiscale Entropy function used in
\citep{starck:pan96} (we call it LOG-MSE in the following) is defined
by:
\begin{eqnarray}
h(w_{j,k,l}) = \frac{\sigma_j}{\sigma_X^2} [w_{j,k,l} - m_j - 
       \mid w_{j,k,l} \mid \log( \frac{\mid w_{j,k,l} \mid}{K_m \sigma_j})]
\label{eqn_logmse}
\end{eqnarray}
where $\sigma_X$ is the total noise standard deviation of the data and
$\sigma_j$ is the noise standard deviation at scale $j$. $K_m$ is a
user-supplied parameter.   
\item{ENERGY-MSE:} The entropy can be defined as the function of the
square of the wavelet coefficients
\citep{starck:sta01_1}:
\begin{eqnarray}
h(w_{j,k,l}) = \frac{w_{j,k,l}^2}{\sigma_j^2} 
\label{eqn_enmse}
\end{eqnarray}
The same multiscale entropy function was also derived in \citet{entropy:hobson04}.
\item{NOISE-MSE:} In \citet{starck:sta01_1}, the entropy is derived
using a modeling of the noise contained in the data:
\begin{eqnarray}
h(w_{j,k,l}) =  \int_{0}^{\mid w_{j,k,l} \mid } P_n(\mid w_{j,k,l} \mid - u) 
   (\frac{\partial h(x)}{\partial x})_{x=u} du
\end{eqnarray}
where $P_n(w_{j,k,l})$ is the probability that the coefficient
$w_{j,k,l}$ can be due to the noise: $ P_n(w_{j,k,l}) =
\mathrm{Prob}(W > \mid w_{j,k,l} \mid) $.
For Gaussian noise, we have:
\begin{eqnarray}
 P_n(w_{j,k,l}) & =  & \frac{2}{\sqrt{2 \pi} \sigma_j} 
 \int_{\mid w_{j,k,l} \mid}^{+\infty} \exp(-W^2/2\sigma^2_j) dW \nonumber \\ 
 & = & \mbox{erfc}(\frac{\mid w_{j,k,l} \mid }{\sqrt{2}\sigma_j})
\end{eqnarray}
and
\begin{eqnarray}
h(w_{j,k,l}) = \frac{1}{\sigma_j^2} \int_{0}^{\mid w_{j,k,l} \mid} u 
              \mbox{ erfc}(\frac{\mid w_{j,k,l} \mid -u}{\sqrt{2} \sigma_j}) du 
\end{eqnarray}
\end{itemize}

{
LOG-MSE presents an indeterminacy when the wavelet coefficient is equal or
close to 0 and the model used in equation~(\ref{eqn_logmse}) is
somewhat ad hoc. This point was raised in \citet{entropy:hobson04}.
A better choice for the LOG-MSE would be
the Herbert and Leaby function \citep{markov:hebert89} (see also the discussion 
in section~\ref{sec_discuss_wave}):
\begin{eqnarray}
h(w_{j,k,l}) \propto  \log \left( 1 +  \frac{\mid w_{j,k,l} \mid}{\sigma_j} \right)
\end{eqnarray}
The ENERGY-MSE is quadratic 
and leads to a strong penalization even for wavelet
coefficients with high signal-to-noise ratio.
Such penalization terms are known to oversmooth the strongest
peaks and should not be used for the weak lensing mass reconstruction. 
The NOISE-MSE is very close to the $l_1$ norm
(i.e. absolute value of the wavelet coefficient) when the coefficient
value is large, which is known to produce good results for the
analysis of piecewise smooth images \citep{cur:elad03}.  We therefore
choose the NOISE-MSE entropy as the most appropriate for the weak
lensing reconstruction problem. Fig.~\ref{fig_pen_mem} 
shows the $l_1$ norm penalization function, the ENERGY-MSE and NOISE-MSE. 
The NOISE-MSE penalization presents a quadratic behavior for small coefficients
and a linear one for larger coefficients. More details are given in section~\ref{sec_discuss_wave}.
}

\subsection{Significant Wavelet Coefficients using the FDR}
In \citet{starck:pan96}, it has been suggested to not apply the
regularization on wavelet coefficients which are clearly detected
(i.e. significant wavelet coefficients). The new Multiscale Entropy
is:
\begin{eqnarray}
h_n(w_{j,k,l}) =  {\bar M}(j,k,l)  h(w_{j,k,l})  
\end{eqnarray}
where ${\bar M}(j,k,l) = 1 - M(j,k,l)$, and $M$ is the multiresolution support  \citep{starck:mur95_2}:
\begin{eqnarray} 
M(j,k,l) = \left\{ \begin{array}{ll} \mbox{ 1 } & 
\mbox{ if
}   w_{j,k,l} \mbox{ is significant} \\ \mbox{ 0 } & \mbox{ if }  
w_{j,k,l}
\mbox{ is not significant} \end{array} \right. 
\end{eqnarray} 
This describes, in a Boolean way, whether the data contains
information at a given scale $j$ and at a given position $(k,l)$.
Commonly, $w_{j,k,l}$ is said to be significant if the probability
that the wavelet coefficient is due to noise is small, i.e. if $P(\mid
W > w_{j,k,l} \mid ) < \epsilon$, where $P$ is a given noise
distribution function.  In the case of Gaussian noise, this amount to
state that $w_{j,k,l}$ is significant if $\mid w_{j,k,l} \mid > k
\sigma_j$, where $\sigma_j$ is the noise standard deviation at scale
$j$, and $k$ is a constant, generally taken between 3 and 5
\citep{starck:mur95_2}.  With this definition, the number of false
detections depends on both the $\epsilon$ value and the image size.

An alternative approach to this detection strategy is the False
Discovery Rate method (FDR) \citep{benjamini95}.  This technique has
recently be introduced for astronomical data analysis
\citep{miller01,hopkins02}.  It allows us to control the average
fraction of false detections made over the total number of
detections. It also offers an effective way to select an adaptive
threshold.  The FDR is given by the ratio :
\begin{eqnarray} 
FDR = \frac{V_{ia}}{D_a}
\end{eqnarray}
where $V_{ia}$ are the number of pixels truly inactive declared active
and $D_a$ are the number of pixels declared active.

This procedure controlling the FDR specifies a rate $\alpha$ between 0
and 1 and ensures that, {\it on average}, the FDR is no bigger than
$\alpha$:
\begin{eqnarray} 
E(FDR) \leq \frac{T_i}{V}.\alpha \leq \alpha
\end{eqnarray}
The unknown factor $\frac{T_i}{V}$ is the proportion of truly inactive
pixels.  A complete description of the FDR method can be found in
\cite{miller01}.  In \citet{hopkins02}, it has been shown that the FDR
outperforms standard methods for sources detection.

{
Here, we use the FDR method; at each resolution level $j$ of the decomposition.
We derive a detection threshold $T_j$ (from a $\alpha_j$ value). We have chosen 
to take a different $\alpha$ value per scale. To fix a $\alpha$ value per scale, 
we used The Receiver Operating Characteristic (ROC) \citep{genovese97} curves 
in order to quantify the quality of the detection at a given scale for different $\alpha$ values. 
We found that the $\alpha_j$ value must increase with scale following the
relation:
$\alpha_j=\alpha_0*2^j$ for the spatial observations and $
\alpha_j=\alpha_0*(1.7)^j$ for the ground observations where $\alpha_0=0.0125$.
We then consider a wavelet coefficient. $w_{j,k,l}$ as significant if its absolute value is larger than $T_j$.}

\subsection{Multiscale Entropy Restoration}
Assuming Gaussian noise, the Multiscale Entropy restoration method
lead to the minimization of the functional,
\begin{eqnarray}
J({\tilde \kappa})= \frac{\parallel {\tilde \kappa}_l^{(E)}  - \tilde \kappa \parallel ^2}
  {2\sigma_n^2} 
  + \beta \sum_{j=1}^{J} \sum_{k,l} h_n( ({\cal W} {\tilde \kappa})_{j,k,l})    
\end{eqnarray}
where $\sigma_n$ the noise standard deviation in \tkle, $J$ the number
of scales,  $\beta$ is the regularization parameter and $\cal W$ is the Wavelet Transform operator.
{
The $\beta$ parameter is calculated automatically
under the constraint that the residual should have a standard deviation 
equal to the noise standard deviation.}
Full details of the minimization 
algorithm can be found in \citet{starck:sta01_1}, as well as the way to determine automatically
the regularization parameter $\beta$.  

{
\subsection{Related Work}
\label{sec_discuss_wave}

\subsubsection{The Generalized Wavelet Regularization}

Using a prior such that a pixel value is a function of its neighborhood
(see \citet{rest:molina01} for more details on the Markov Random Field model), 
the Bayesian solution consists in adding the following penalization on the solution:
\begin{eqnarray}
{\cal C} ({\tilde \kappa}) & = & \beta \sum_{x}\sum_{y} 
 \left( \phi({\tilde \kappa}(x,y) - {\tilde \kappa}(x,y+1))^2 \right.
 +  \nonumber \\
  &  & + \left. \phi({\tilde \kappa}(x,y) - {\tilde \kappa}(x+1,y))^2  
 \right)^{\frac{1}{2}} \nonumber 
\label{edgepres}
\end{eqnarray}
The function $\phi$, called {\em potential function},
 is an edge preserving function. The term $\beta
\sum_{x}\sum_{y} \phi(\parallel \nabla I \parallel (x,y))$
 can also be interpreted as the Gibbs energy of a Markov Random Field.
Generally, functions $\phi$ are chosen with a quadratic part which ensures
a good smoothing of small gradients \citep{markov:green90},
and a linear behavior which cancels the penalization of large gradients \citep{markov:bouman93}:
\begin{enumerate}
\item $\lim_{t \rightarrow 0} \frac{\phi^{'}(t)}{2t}   =  1$, smooth faint gradients.
\item $ \lim_{t \rightarrow \infty} \frac{\phi^{'}(t)}{2t}  =   0$,  preserve strong gradients.
\item $  \frac{\phi^{'}(t)}{2t} $ is strictly decreasing.
\end{enumerate}
Such functions are often called $L_2$-$L_1$ functions. Examples of $\phi$ functions:
\begin{enumerate}
\baselineskip=0.2truecm
\itemsep=0.01truecm
\item $\phi_{q}(x)  = x^2 $:  quadratic function. 
\item $\phi_{TV}(x)  = \mid x \mid$: Total Variation. 
\item $\phi_2(x) = 2\sqrt{1+x^2}-2$:  Hyper-Surface \citep{markov:charbonnier97}.
\item $\phi_3(x) = x^2/(1+x^2)$  \citep{markov:geman85}.    
\item $\phi_4(x) = 1-e^{-x^2}$ \citep{markov:perona90}.
\item $\phi_5(x) = \log(1+x^2)$ \citep{markov:hebert89}. 
\end{enumerate}

\begin{figure*}[htb]
\centerline{
\vbox{
\hbox{
}
}}
\caption{Examples of potential function $\phi$.}  
\label{potential_phi}
\end{figure*}
Figure~\ref{potential_phi} shows different $\phi$ functions.

It has been shown that this concept can be generalized in the wavelet domain, leading  
to a multiscale wavelet penalization term \citep{markov:jalobeanu01}:
  \begin{eqnarray}
\label{sect_wt_regul}
  {\cal C}_w({\tilde \kappa})  =  
  \beta \sum_{j,k,l} \phi(\parallel ({\wave} {\tilde \kappa})_{j,k,l} \parallel_p)  
\end{eqnarray}
When $\phi(x) = x$ and $p=1$, it corresponds to the $l_1$ norm of the 
wavelet coefficients. In this framework, the multiscale entropy deconvolution
method is only one special case of the wavelet constraint deconvolution method. 

\begin{figure}[htb]
\centerline{
\hbox{
}}
\caption{Penalization functions: dashed, $l_1$ norm 
(i.e. ${\phi}(w) = \mid w \mid)$; dotted $l_2$ norm ${\phi}(w) = \frac{w^2}{2}$ (i.e. ENERGY-MSE);
continuous, multiscale entropy function (NOISE-MSE).}
\label{fig_pen_mem}
\end{figure}
 Figure~\ref{fig_pen_mem} shows
the multiscale entropy penalization function. The dashed line corresponds to a $l_1$ penalization 
(i.e. ${\phi}(w) = \mid w \mid)$, the  dotted line to a 
$l_2$ penalization ${\phi}(w) = \frac{w^2}{2}$, and the continuous line to the
 multiscale entropy function. We can immediately see that the multiscale entropy function 
 presents a quadratic behavior for small values, and is closer to the $l_1$ penalization 
 function for large values. Penalization function with a $l_2$-$l_1$ behavior are known to be a good 
 choice for image restoration.

\subsubsection{Multiscale MEM and ICF}
The multichannel ICF-MEM method \citep{entropy:weir91,entropy:weir92} consists in assuming that  the 
visible-space image $O$ is formed by a weighted sum of 
the visible-space image channels $O_j$,  $O = \sum_{j=1}^{N_c} p_j  O_j$
where $N_c$ is the number of channels and $O_j$ is the result of the
convolution between a hidden image $h_j$ with a low-pass filter (ICF) $C_j$,
called ICF (Intrinsic Correlation Function) (i.e. $O_j= C_j * h_j$).
In practice, the ICF is a Gaussian.
The MEM-ICF constraint is:
\begin{eqnarray}
  {\cal C}_{ICF} = \sum_{j=1}^{N_c} \mid h_j \mid - m_j
      - \mid h_j \mid \log \left( \frac{ \mid h_j \mid}{m_j}\right)
\end{eqnarray}
In \citet{entropy:hobson04}, it was argued that the multiscale entropy  
is merely a special case of the intrinsic correlation function approach, where 
we replace the ICF kernel by a wavelet function. 
From the strict mathematical point of view, this is right, but this vision
minimizes completely the improvement related to the wavelets.
All the concepts of sparse representation (which is the key of the wavelet success in many applications), 
fast decomposition and reconstruction, zero mean coefficients (which allows us to get
wavelet coefficients which are independent of the background 
and to derive a robust noise modeling) do not exist in the ICF-MEM approach.
Furthermore, ICF-MEM approach requires to estimate accurately the background, which  
may be sometimes a very difficult task, and it has be shown \citep{entropy:bontekoe94} 
that the solution depends strongly on this estimation. On the contrary, Multiscale MEM needs only
an estimation of the noise standard deviation, which is easy to determine. 

For all these reasons, we prefer to keep our vision of the multiscale entropy method 
as a specific case of the generalized  wavelet regularization techniques rather than as
an extension of the ICF approach.
}

\section{Results}
\label{sect_results}

\subsection{Comparison of methods}

\begin{table*}
\begin{center}
\begin{tabular}{c|c|c}
\hline
  Method       & Error ($n_g = 20$ $gal/amin^2$) &  Error ($n_g = 100$ $gal/amin^2$)\\
\hline \hline
Gaussian Filtering ($\sigma_G = 1$ amin) & 1.108 & 0.775\\
Gaussian Filtering ($\sigma_G = 2.5$ amin) & 0.9138 & 0.868\\
Wiener Filtering                   & 0.888 & 0.770\\
MEM-LensEnt2 & 1.091 & 0.821\\
Multiscale Entropy Filtering       & 0.888 & 0.746 \\
\hline
\end{tabular}
\end{center}
\caption{Standard deviation of the reconstruction error with five different methods.}
\label{tab_mse}
\end{table*}

\begin{figure*}[htb]
\centerline{
\hbox{
\hspace{0.2cm}
}}
\caption{Standard deviation versus scale for the ground-based simulation (left) and the
space-based simulation (right).}
\label{fig_res_recons}
\end{figure*}

\begin{figure*}[htb]
\centerline{
\hbox{
\hspace{0.2cm}
}}
\caption{Log Power Spectrum of the Error by Multiscale Entropy Filter and MEM  for the ground-based simulation (left) and the space-based simulation (right).}
\label{powerspec}
\end{figure*}

\begin{figure*}[htb]
\vbox{
\centerline{
\hbox{
\hspace{0.2cm}
}}
\centerline{
\hbox{
\hspace{0.2cm}
}}
\centerline{
\hbox{
\hspace{0.2cm}
}}
}
\caption{Restoration of the  $2\times2$ square degrees ground-based observation (left) and spatial  observation (right).
From top to bottom, Gaussian filtering, Wiener filtering and Multiscale Entropy filtering.}
\label{fig_rec_kappa}
\end{figure*}

We have used a simulated data set obtained using a standard
$\Lambda$-CDM cosmological model.  A part of the $\kappa$ mass map and
the shear maps is shown in Fig.~\ref{kb0}.  The field size is
$2\times2$ square degrees, sampled with $1024*1024$ pixels.  
 
Noisy shear maps, corresponding to both spatial (i.e. $n_g = 100$ gals amin$^{-2}$)
and ground-based observations (i.e. $n_g = 20$ gals amin$^{-2}$), are
created using equation~\ref{eq_gamma}.  Then we have reconstructed the
two noisy mass maps from equation~\ref{eqn_reckE} and applied the
following methods:
\begin{enumerate}
\item Gaussian filtering with a standard deviation equal to $\sigma_G=1$ amin. 
\item Gaussian filtering with a standard deviation equal to $\sigma_G=2.5$ 
 amin. 
\item Wiener filtering.
\item Maximum Entropy Method (MEM) using the {\bf LensEnt2} package.
As this code has not been designed for manipulating large images, we had
to restrict the restoration by this method to a field size 
of$ 0.5 \times 0.5$ square degree,  sampled with 256*256 pixels.
Since the {\bf LensEnt2} maps are positivity constrained, as recommended by the author of 
the {\bf LensEnt2} package, we have recovered a physical mass 
by transforming the outputs such that the minimum 
convergence in the central quarter of the reconstruction is zero.
To optimize the ICF, we have maximized the Bayesian evidence value as a function of ICF width,
and found that maximum evidence is around 210 arcsec for ground observations 
and around 180 arcsec for space observations.

\item Multiscale Entropy method.
\end{enumerate}

The evaluation is done by i) visual inspection of the images, ii)
calculating the standard deviation between the original $\kappa$ mass
map and the reconstructed map (i.e. $E = \frac{STD(\kappa - \tilde
\kappa)}{STD(\kappa)}$), iii) calculating the standard deviation for
each of their wavelet scales (i.e. $E_j = \frac{STD(({\cal W}
\kappa)_j - ({\cal W} \tilde \kappa)_j)}{STD(({\cal W} \kappa)_j)}$) and iv) 
calculation the power spectrum of the error $E$ (for MEM and multiscale entropy methods).

The values $\sigma_G = 1$ amin and $\sigma_G = 2.5$ amin have been
chosen to optimize the Gaussian filtering for $n_g = 100$ gals
arcmin$^{-2}$ and $n_g = 20$ gals arcmin$^{-2}$, respectively.
Table~\ref{tab_mse} gives the standard deviation of the error for the
four reconstructed mass maps.  It shows that i) the Wiener filtering is
better than the Gaussian filtering and the MEM-LensEnt2 method and ii)
the Multiscale Entropy outperforms the three other methods.
 
Fig.~\ref{fig_res_recons} shows the error versus the scale (each wavelet scale) for both
simulations using the Gaussian filtering (continuous line), the Wiener
filtering (dotted line), the MEM-LensEnt2 filtering (dashed line) and the Multiscale Entropy
filtering (dotted-dashed line). The wavelet scales 1 to 6 correspond to scales
of $0.12, 0.23, 0.47, 0.94, 1.87, 3.75$ amin respectively. 
We can see that the Multiscale Entropy method produces better results for all scales.

Fig.~\ref{powerspec} shows the log power spectrum of the error. 
It is very consistent with the previous one.
Indeed, the MEM error becomes very important toward the smallest frequencies (largest wavelet scales).
The same experiment has been done with a smallest ICF (ICF=120 for the spatial simulation), but
the result is worse, which is not surprising since the ICF value was chosen to get the best results.

Fig.~\ref{fig_rec_kappa} shows from top to bottom
the reconstructed maps for the Gaussian, the Wiener and Multiscale Entropy filtering.
Fig.~\ref{fig_rec_kappa} left corresponds to ground-based observations (i.e. $n_g=20$) and
Fig.~\ref{fig_rec_kappa} right corresponds to spatial  observations (i.e. $n_g=100$).

Fig.~\ref{kappa_MEM} shows the denoising results on a portion of the 
previous image. Fig.~\ref{kappa_MEM}  shows the original noise free simulated image 
of the $0.5\times0.5$ square degrees field (upper left), the Multiscale Entropy Filtering
for the spatial simulated observations ($n_G=100$) (upper right), the MEM-LensEnt2 restoration
for the ground based observations (bottom left) and the spatial observations (bottom right). 

The computation time for the 1024*1024 pixels map is 4 minutes 
for the Multiscale Entropy method, 26 seconds for the Wiener filtering  
and 4 seconds for the Gaussian smoothing. 
The computation time for the 256*256 pixels map is around 60 minutes
(it depends on the convergence of the result) using the MEM-LensEnt2 package.

\begin{figure*}[htb]
\vbox{
\centerline{
\hbox{
\hspace{0.2cm}
}}
\vspace{0.3cm}
\centerline{
\hbox{
\hspace{0.2cm}
}}
}
\caption{In a region of $0.5\times0.5$ square degrees, a sixteenth of the original field : Upper left,  simulated mass map, upper right, Multiscale entropy filtering for $n_g=100$ gal/arcmin$^{2}$. Bottom left, MEM filtering for $n_g=20$ gals/amin$^{-2}$ ($ICFwidth=210$) and bottom right for $n_g=100$ gals/amin$^{-2}$ ($ICFwidth=180$).}
\label{kappa_MEM}
\end{figure*}

\subsection{Robustness to missing data}

\begin{figure*}[htb]
\vbox{
\centerline{
\hbox{
\hspace{0.2cm}
}}
\vspace{0.3cm}
\centerline{
\hbox{
\hspace{0.2cm}
}}
}
\caption{Upper left, noisy shear map ($n_g=100 gal/arcmin^{2}$). Upper right, Gaussian filtering. Bottom left, Wiener 
filtering, and bottom right, Multiscale Entropy filtering.}
\label{kappa_rec_mem_miss}
\end{figure*}

During the observations, various problem can cause a loss of data in
the image.  For example, it can be due to a defect of the camera CCD,
generating a dark line or a dark row in the image, or to the presence
of a bright star in the field of view which forces us to remove part
of the image.  In order to study this problem, we mask two rectangular
areas, setting all pixel values to 0, in the shear maps $\gamma_1$ and
$\gamma_2$. By inverse filtering, we have derived the noisy mass map
$\kappa_l$ in which we can also visualize the lack of data
(Fig.~\ref{kappa_rec_mem_miss} upper left).  Then we have applied the
three methods, Gaussian filtering, Wiener filtering and Multiscale
Entropy, to the noisy mass map and the results can be seen
respectively in Fig.~\ref{kappa_rec_mem_miss} upper right,
Fig.~\ref{kappa_rec_mem_miss} bottom left and bottom right.  We can
see that all three methods are robust to the missing data. Note
however that, for the Wiener filtering, we have assumed perfect
knowledge of the power spectrum of $\kappa$, while, in practice, its
estimation is made more complicated by the complex field geometry.

\begin{figure*}[htb]
\centerline{
\hbox{
\hspace{0.2cm}
}}
\caption{Standard deviation versus scale for the ground-based simulation(left) and the spatial simulation (right) with missing data }
\label{fig_res_recons_missing}
\end{figure*}

Fig.~\ref{fig_res_recons_missing} shows the error versus the scale for both
simulations using the Gaussian filtering (continuous line), the Wiener
filtering (dotted line) and Multiscale Entropy.  We can see that the
Multiscale Entropy still produces better results at all scales.  
Bayesian methods such MEM could also take into account properly missing data, however not
in a straightforward way as when using wavelets.

\subsection{Cluster detection}

\begin{figure*}[htb]
\centerline{
\hbox{
\hspace{0.2cm}
}}
\caption{Isophotes of the detected wavelet coefficients for the space-based simulation overplotted on
the original mass map: left, using the $k\sigma$ standard approach 
(with $k=3,4,5$) and
right using the FDR method.}
\label{fig_cmp_fdr}
\end{figure*}

\begin{figure*}[htb]
\centerline{
\hbox{
\hspace{0.2cm}
}}
\caption{Zoom of the previous maps}
\label{fig_cmp_fdr_zoom}
\end{figure*}

\begin{figure*}[htp]
\vbox{
\centerline{
\hbox{
\hspace{0.2cm}
}}
\vspace{0.3cm}
\centerline{
\hbox{
}}
}
\caption{The isophotes represent the detected clusters using 
the Gaussian filtering (upper left), the Wiener filtering (upper right)
and the wavelet-FDR method.}
\label{isogw}
\end{figure*}

Another important aspect of the weak shear mass reconstruction is the
possibility to detect clusters and to build a catalog. Here, using the
FDR in the wavelet space, we detect as significant a set of wavelet
coefficients. We built an isophote map, where each isophote level
corresponds to the detection level in a given scale. This isophote is
overplotted on the true mass map, which allows us to visually check
the false detections and the missed detections. A cluster surrounded
by two isophotes means that it has been detected at two
scales. Fig.~\ref{fig_cmp_fdr} left shows the isophote map when we use
the regular $k\sigma$ thresholding and Fig.~\ref{fig_cmp_fdr} left
right shows the isophote map when we use the FDR method. We see that
the FDR is more sensitive than the $k\sigma$ method for the detection,
 without being contaminated
by a large number of false detections.  Fig.~\ref{fig_cmp_fdr_zoom}
shows a zoom of these two maps.
Figure~\ref{isogw} shows a comparison between the Gaussian filtering, the Wiener
filtering and FDR-Wavelet method for the detection of clusters. In the Gaussian 
and Wiener maps, the isophotes corresponds to a $k\sigma$ detection level where $k=3,4,5$.
It shows clearly how the FDR-Wavelet method outperforms the other methods.

\subsection{E/B Decomposition}
As explained in \S\ref{eb}, a simple diagnostic test for a wide range
of systematic effects is to search for the presence of B-mode in the
lensing maps.  In order to test it, we have simulated mass maps with a B-mode.

\begin{figure*}[htb]
\centerline{
\hbox{
\hspace{0.2cm}
}}
\caption{left mass map (E-mode), right mass map (B-mode)}
\label{simEB}
\end{figure*}

\begin{figure}[htb]
\centerline{
\hbox{
}}
\caption{Noisy simulated mass map}
\label{noisy_simEB}
\end{figure}

\begin{figure*}[htb]
\centerline{
\hbox{
\hspace{0.2cm}
}}
\caption{left filtered noisy mass map (E-mode), right filtered noisy mass map (B-mode)}
\label{fdr_noisy_simEB}
\end{figure*}

Fig.~\ref{simEB} left shows a simulated mass map with a lensing E-mode
signal (left) and an arbitrary B-mode signal (left). As usual, we have
added a realistic space-based Gaussian noise to the shear of this
simulation.  Fig.~\ref{noisy_simEB} shows the noisy mass map
resulting. Using the Multiscale Entropy filtering, we have then
reconstructed the two components of the mass map (see \S\ref{eb}):
E-mode in Fig.~\ref{fdr_noisy_simEB} left and B-mode in
Fig.~\ref{fdr_noisy_simEB} right.  We see clearly that the wavelet
separation of the E and B modes is very good.  Indeed, the two main
features in the B-mode have well been recovered, without interfering
with the reconstruction of the E-mode.

\section{Conclusion}
\label{sect_conclusion}

We have presented in this paper a new way to reconstruct weak lensing
mass maps. We have modified the Multiscale Entropy method in order to
take into account the FDR.  We have shown that this new method
outperforms several standard techniques currently used for the weak
shear mass reconstruction. The visual aspect as well as objective
criteria, such the rms of the error or the rms per scale of the error,
clearly show the advantages of the proposed approach.  Experiments
have demonstrated that it is also robust to missing data. We have also
shown that a E/B mode separation can also be performed using this
method, thus providing a useful test for the spatial distribution of
residual systematics. Our method allows us also to build a catalog of
clusters and the use of FDR leads to a clear improvement in
sensitivity, compared to what has been done previously with wavelets.

\subsubsection*{Software}
The software related to this paper, {\bf MR/Lens}, and its full documentation 
are available from the following web page:
\begin{verbatim}
       http://jstarck.free.fr/mrlens.hmtl 
 \end{verbatim}
\begin{acknowledgements}
We wish to thank Savita Mathur for her initial work on the subject,
 Richard Massey and Bedros Afeyan for useful discussions.
We would like to particularly acknowledge Phil Marshall 
for his help relative to the use of the LensEnt2 package.
\end{acknowledgements}


\end{document}